\documentclass[aps,twocolumn,pra,superscriptaddress]{revtex4}
\usepackage{epsfig,graphicx,times}
\usepackage{amstext}
\usepackage{amsmath}
\usepackage{amsfonts}
\usepackage{amssymb}
\usepackage{graphicx}
\usepackage{latexsym}
\usepackage{subfigure}
\usepackage{bm}
\usepackage{bbm}
\usepackage[colorlinks,citecolor=blue,linkcolor=blue,hyperindex,CJKbookmarks,dvipdfm]{hyperref}

\begin{document}
\title{Parameter estimation with limited access of measurements}
\author{Jianning Li}
\affiliation{Center for Quantum Sciences and School of Physics, Northeast Normal University, Changchun 130024, China}
\author{Dianzhen Cui}
\affiliation{Center for Quantum Sciences and School of Physics, Northeast Normal University, Changchun 130024, China}
\author{X. X. Yi\footnote{yixx@nenu.edu.cn}}
\affiliation{Center for Quantum Sciences and School of Physics, Northeast Normal University, Changchun 130024, China}
\affiliation{Center for Advanced Optoelectronic Functional Materials Research, and Key Laboratory for UV Light-Emitting Materials and Technology of Ministry of Education, Northeast Normal University, Changchun 130024, China}

\date{\today}

\begin{abstract}
Quantum parameter estimation holds the promise of quantum technologies, in which physical parameters can be measured with much greater precision than what is achieved with classical technologies. However, how to obtain a best precision when the optimal measurement is not accessible is still an open problem. In this work, we present a theoretical framework to explore the parameter estimation with limited access of measurements by analyzing the effect of non-optimal measurement on the estimation precision. We define a quantity $\Lambda$ to characterize the effect and illustrate how to optimize observables to attain a bound with limited accessibility of observables. On the other side, we introduce the minimum Euclidean distance to quantify the difference between an observable and the optimal ones in terms of Frobenius norm and find that the measurement with a shorter distance to the optimal ones benefits the estimation. Two examples are presented to show our theory. In the first, we analyze the effect of non-optimal measurement on the estimation precision of the transition frequency for a driven qubit. While in the second example, we consider a bipartite system, in which one of them is measurement inaccessible. To be specific, we take a toy model, the NV-center in diamond as the bipartite system, where the NV-center electronic spin interacts with a single nucleus via the dipole-dipole interaction. We achieve a precise estimation for the nuclear Larmor frequency by optimizing only the observables of the electronic spin. In these two examples, the minimum Euclidean distance between an observable and the optimal ones is analyzed and the results show that the observable closed to the optimal ones better the estimation precision. This work establishes a relation between the estimation precision and the distance of the non-optimal observable to the optimal ones, which would be helpful for experiment to seek the best observable in parameter estimation.
\end{abstract}
\maketitle

\section{Introduction}
Quantum metrology is advanced by quantum mechanics, which concerns the estimation of unknown physical parameters \cite{Helstrom1967,Helstrom1976,Holevo1982} and aims at improving the estimation precision beyond classical limit. In quantum metrology, the estimation precision is bounded by the quantum Cram\'{e}r-Rao bound (QCRB) \cite{Cramer1946,Braunstein1994,Braunstein1996}, which states that the variance of the estimation is at least as high as the inverse of the quantum Fisher information (QFI).

The general process of the parameter estimation includes the following three steps: preparation, parametrization and measurement. All the steps need to be optimized to obtain a precise estimation. In the step of preparation, one could improve the precision employing quantum features, such as squeezing \cite{Maccone2020} and entanglement \cite{Giovannetti2004,Giovannetti2006,Nagata2007}. For example, the entangled state with optimal measurements could improve the precision of parameter estimation with respect to classical methods \cite{Giovannetti2006}. The step of parametrization in general is performed by the time evolution governed by a Hamiltonian  \cite{Boixo2007,Toth2014,Pang2014,Pang2017,Elias2017}, and a better precision could be obtained by Hamiltonian extensions or subtractions \cite{Elias2017}. In the step of measurement, an optimization of the measurement is needed to obtain the highest estimation precision bounded by the QCRB \cite{Braunstein1994,Braunstein1996}.

The Fisher information in the case of single-parameter estimation is the key quantity representing the ultimate estimation precision of the unknown parameters and that the QFI is given by the symmetric logarithmic derivative (SLD), the measurement to saturate the QCRB is the projective measurement on the eigenvectors of the SLD \cite{Helstrom1976,Braunstein1994,Braunstein1996}. In practice, however, the realization of the optimal measurement might be a challenging task due to the following reasons: (1) The measurements performed are non-optimal due to the technological difficulties or unavoidable noises in realistic quantum measurement \cite{Davis2022,Zhai2023,Lee2023,Kolodynski2010,Demkowicz2012,Escher2011,Nichols2016,Takeuchi2007,Clerk2010,Chen2019,Riccardi2022,Kurdzialek2023} and (2) the optimal measurement may be inaccessible due to the limited accessibility of measurement in the systems \cite{Montenegro20221}, for example, it is difficult to measure the nuclear Larmor frequency directly in the coupled nucleus-electron system \cite{Jelezko2006,Lee2013,Shi2014}. With this consideration, one has to resort to partial accessibility of measurements and acquire the information about the whole system via its compositing subsystems \cite{Mishra2021,Montenegro20222}. Indeed some works reported the estimation precision obtained by local control \cite{Kiukas2017} or local measurement \cite{Troiani2016,Li2023}, and they found that the precision by this partial measurement is no better than the optimal global measurement \cite{Li2023,Paris2009,Yuan2017}.

Taking the accessability of optimal measurements into account, one may wonder how the non-optimal measurements affect the ultimate estimation precision? And if all measurements in the whole system are not accessible, how to optimize the observables of the subsystem to improve the estimation precision? What is the bound in this situation? Before answering the above questions, we should point out that if one is only concerned about which measurements are better for the estimation with limited access of measurements, there are not only one criterion to do this, such as minimizing the mean-square error or maximizing the classical Fisher information. However, those comparisons between different measurements is rough because it omits the important information about how the deviation of the non-optimal measurements from the optimal ones generates the effect on the low bound of the estimation. Therefore, in the present paper, we solve these questions with a general theoretical framework to analyze the effect of non-optimal measurement on the estimation precision. We define a non-negative quantity $\Lambda$ to characterize the effect such that an observable with a small $\Lambda$ leads to the estimation with high precision. Furthermore, we demonstrate how to optimize the observables and give a bound that can be attained with limited accessibility of measurements. On the other hand, we introduce the minimum Euclidean distance in terms of the Frobenius norm \cite{Horn2012} to quantify the distance between the performed measurements and the optimal ones. We find that the measurement with a shorter distance to the optimal ones benefits the estimation precision. The theoretical framework is then applied to a driven qubit and a toy model, the NV-center in diamond, a coupled nucleus-electron bipartite system. In the driven qubit, we show how the non-optimal measurement affects the estimation precision of its transition frequency, while in the coupled nucleus-electron bipartite system, we demonstrate the optimization of the electron observables to estimate the nuclear Larmor frequency. An estimation precision by optimizing only local observables \cite{Ariano2003,Hotta2004} of the electron is given and the effect of the noise on the estimation precision with different local observables is also given. In addition, an estimation precision with the optimizations of joint observables \cite{Busch1986} is discussed. Finally, we compare the precision obtained from different accessibility of the measurements. As expected, the precision obtained by the partial accessibility is no better than that by whole accessibility of the measurements. In these two examples, we also compare the minimum Euclidean distance between an observable and the optimal ones with the non-negative quantity $\Lambda$, and conclude that the observable closed to the optimal ones better the estimation precision.

The remainder of this paper is organized as following. We present a theoretical framework to analyze the effect of the non-optimal observables on the estimation precision, and apply the framework to composite systems with limited accessibility of measurement in Sec.~\ref{sec1}. In Sec.~\ref{sec2}, we demonstrate the theory with a driven qubit and the NV-center in diamond, where we focus on the estimation of transition frequency and the nuclear Larmor frequency. The minimum Euclidean distance between the non-optimal observables and the optimal ones is calculated and discussed. Finally, we conclude and discuss the results in Sec.~\ref{sec3}. The appendices are provided as supplemental materials for the discussions in the main text.

\section{Framework}\label{sec1}
The theory of quantum metrology states that the estimation precision of an unknown parameter depends on the measurement of an observable. This suggests that in order to obtain a better estimation precision, one has to optimize the measurement, namely, to choose an suitable observable to measure. In this work we will use the notation of measurement and observable alternatively when there is no confusion. As aforementioned, the measurement performed for the estimation of unknown parameters is generally non-optimal. In this section, we will first present a theoretical framework to analyze the effect of the non-optimal observables on the precision, then we apply the framework into composite systems to optimize the observables with limited accessibility of measurements. Finally, we introduce the minimum Euclidean distance to quantify the distance between the performed measurement and the optimal ones.

\subsection{The effect of non-optimal measurement on the estimation precision and the optimization with limited accessibility of measurements}
Theoretically, for a parameterized density matrix $\rho_\theta$, where $\theta$ is the unknown parameter to be estimated, the variance of the estimation is lower bounded by the QCRB \cite{Cramer1946,Braunstein1994,Braunstein1996}
\begin{eqnarray}
\begin{aligned}
(\delta\hat{\theta})^2\geq\frac{1}{\mathcal{F}_Q},
\label{QCR_ieq2}
\end{aligned}
\end{eqnarray}
where $\mathcal{F}_Q=\textrm{Tr}(\rho_\theta L_\theta^2)$ is the QFI and $L_\theta$ is the SLD satisfying $\frac{\partial\rho_\theta}{\partial\theta}=\frac{1}{2}(\rho_\theta L_\theta+L_\theta\rho_\theta)$ \cite{Braunstein1994,Braunstein1996}. One sufficient condition of attaining the low bound is that the optimal observable $A_{opt}$ is commutative with $L_\theta$, i.e., $[A_{opt},L_\theta]=0$ (the proof is in Appendix \ref{appendixA}). Consider a non-optimal estimation where the non-optimal observable $A$ is written as
\begin{eqnarray}
\begin{aligned}
A=A_{opt}+\delta A,
\label{non_optimal}
\end{aligned}
\end{eqnarray}
where $\delta A$ is the deviation from the optimal observable $A_{opt}$. Submitting Eq. (\ref{non_optimal}) into the error propagation formula
\begin{eqnarray}
\begin{aligned}
(\delta\theta)^2=\frac{\langle(\Delta A)^2\rangle}{|\partial_{\theta}\langle A\rangle|^2},
\label{non_optimal0}
\end{aligned}
\end{eqnarray}
where $\Delta A=A-\langle A\rangle$ and $\langle A\rangle=\textrm{Tr}(A\rho_\theta)$ is the average over the parameterized density matrix $\rho_\theta$, we obtain
\begin{eqnarray}
\begin{aligned}
(\delta\theta)^2=\frac{1}{\varepsilon\mathcal{F}_Q}+\epsilon,
\label{error_propagation2}
\end{aligned}
\end{eqnarray}
where $\varepsilon=\left|1+\frac{\partial_{\theta}\langle \delta A\rangle}{\partial_{\theta}\langle A_{opt}\rangle}\right|^2$, $\epsilon=\frac{\langle (\delta A)^2\rangle-\langle \delta A\rangle^2+\eta}{|\partial_{\theta}\langle A_{opt}\rangle+\partial_{\theta}\langle \delta A\rangle|^2}$, and $\eta=2\left(\textrm{Re}\langle A_{opt}\delta A\rangle-\langle A_{opt}\rangle\langle\delta A\rangle\right)$. Equation (\ref{error_propagation2}) describes the relation of the low bound in Eq. (\ref{QCR_ieq2}) and estimation precision upon the non-optimal observable $A$. Next, we define $\Lambda=(\delta\theta)^2-\textrm{min}[(\delta\hat{\theta})^2]$ to describe the effect of non-optimal observable on the estimation precision. Straightforward calculation follows,
\begin{eqnarray}
\begin{aligned}
\Lambda=-\frac{1}{\varepsilon\mathcal{F}_Q}\left(\left|\frac{\partial_{\theta}\langle \delta A\rangle}{\partial_{\theta}\langle A_{opt}\rangle}\right|^2+2\frac{\partial_{\theta}\langle \delta A\rangle}{\partial_{\theta}\langle A_{opt}\rangle}\right)+\epsilon.
\label{error_propagation3}
\end{aligned}
\end{eqnarray}
Equation (\ref{error_propagation3}) shows that once one know the deviation $\delta A$, we can obtain the effect of non-optimal observable on the estimation precision directly. In addition, it is easy to prove that $\Lambda$ is always non-negative, i.e., $\Lambda\geq0$. For details of proof, see Appendix \ref{appendixA}. By the definition of $\Lambda$, an observable with smaller $\Lambda$ corresponds to a more precise estimation. Next, we will focus particularly on the optimization of the local observables and joint observables and give the details by taking a bipartite system as an example.

The Hamiltonian describing the composite quantum system considered can be expressed as
\begin{eqnarray}
\begin{aligned}
H(\theta)=H^{a}+H^b+H_I,
\label{general_H}
\end{aligned}
\end{eqnarray}
where $H^i$, $i\in\{a,b\}$, are the free Hamiltonian of the subsystems, $H_I$ describes the interaction between the subsystems. The estimated parameter $\theta$ is encoded into a density matrix $\rho_0$ by the unitary dynamics $\rho_\theta=U\rho_0U^\dagger$ with $U=e^{-iH(\theta)t}$, where $\rho_\theta$ denotes the parameterized density matrix. With the assumption that all measurements are accessible in the subsystems, there are three different kinds of choice for the observables to carry out the optimization: the local observables (i) $A=A^a\otimes \mathbb{I}^b$ and (ii) $A=\mathbb{I}^a\otimes A^b$ correspond to local measurements $\{O^i_k\}$ with outcomes $a^i_k$ and probability $p^i_k(\theta)=\textrm{Tr}(\rho_\theta O^i_k)$, where $\mathbb{I}^i$ and $A^i$ are identity operators and arbitrary observables for different subsystems, $O^i_k$ satisfy $\sum_kO^i_k=\mathbb{I}^i$. (iii) The joint observable, $A=A^a\otimes A^b$, whose outcomes $a^a_k$ and $a^b_l$ for $A^a$ and $A^b$ are coexistent \cite{Busch1986} for different subsystems by performing the joint measurements $\{\mathcal{O}_{kl}\}$, which takes $O^i_{k(l)}=\sum_{l(k)}\mathcal{O}_{kl}$ and the corresponding joint probability is $p_{kl}(\theta)=\textrm{Tr}(\rho_\theta\mathcal{O}_{kl})$ \cite{Yu2010,Andersson2005}.

The above local and joint observables are generally non-optimal so the low bound in Eq. (\ref{QCR_ieq2}) can not be saturated. For observable $A$ that is not optimal and takes the forms in (i), (ii) and (iii), the classical Fisher information \cite{Fisher1922,Fisher1925}
\begin{eqnarray}
\begin{aligned}
F_c(\theta; A)=\sum_j\frac{1}{p_j(\theta)}\left[\frac{\partial p_j(\theta)}{\partial \theta}\right]^2,
\label{FI}
\end{aligned}
\end{eqnarray}
with $p_j(\theta)$ being the probability for local or joint measurements is bounded by \cite{Escher2011} for case (i) and (ii),
\begin{eqnarray}
\begin{aligned}
(\delta\theta)^2\geq\frac{1}{\mathcal{F}_Q^i}\geq\frac{1}{\mathcal{F}_Q},
\label{new_QCRB}
\end{aligned}
\end{eqnarray}
where $\mathcal{F}_Q^a\equiv\textrm{max}F_c^a(\theta; A^a\otimes \mathbb{I}^b)$ and $\mathcal{F}_Q^b\equiv\textrm{max}F_c^b(\theta; \mathbb{I}^a\otimes A^b)$ are the maximum classical Fisher information with optimal local observable in different subsystems, namely, the QFI of different subsystems (the details for calculating subsystem QFI are shown in Appendix \ref{appendixB}). Although inequality (\ref{new_QCRB}) shows the bounds of the variance for different measurement performed on the subsystems, there is no specific relation for which subsystem is better for the estimation. In practice, one should carefully choose the suitable subsystem and observable to obtain a precise estimation as high as possible with limited accessibility of measurements. For scenario (iii), we have the following inequality \cite{Lu2012}
\begin{eqnarray}
\begin{aligned}
(\delta\theta)^2\geq\frac{1}{\mathcal{F}_Q^i}\geq\frac{1}{\mathcal{F}_Q^{a\otimes b}}\geq\frac{1}{\mathcal{F}_Q},
\label{new_QCRB3}
\end{aligned}
\end{eqnarray}
where $\mathcal{F}_Q^{a\otimes b}\equiv\textrm{max}F_c^{a\otimes b}(\theta; A^a\otimes A^b)$ is the maximum classical Fisher information with optimal joint observable. Equation (\ref{new_QCRB3}) shows that the variance of the estimation through the optimal joint observable is better than the local ones, which can be understood as the latter is a subset of the formers so that the estimation precision is no more than the former.

\subsection{Minimum Euclidean distance}
We have given a theoretical framework to analyze the effect of the non-optimal observables on the estimation precision and illustrate how to optimize the observables with limited accessibility of measurements in composite systems in the last subsection. A natural question arises: what is the connection between the estimation precision and the non-optimality of the observables? Here, we introduce the minimum Euclidean distance between the non-optimal observables $A$ and the optimal ones to characterize the non-optimality of the measurement, it is defined by
\begin{eqnarray}
\begin{aligned}
\mathcal{D}=\textrm{min}||A-\mathcal{L_\theta}||=||\delta A||,
\label{distance}
\end{aligned}
\end{eqnarray}
where $\mathcal{L}_\theta \in \{M| [L_\theta,M]=0\}$ are the optimal observables and commutative with SLD $L_\theta$, $||\cdot||=\sqrt{\sum_{j,k}|(\cdot)_{j,k}|^2}$ is the Frobenius norm \cite{Horn2012}. Equation (\ref{distance}) shows how to calculate the distance between two observables in terms of parameter estimation, which possesses the following properties: (a) Non-negative, $\mathcal{D}\geq0$; (b) The observable with a shorter minimum Euclidean distance to the optimal ones is better for the estimation; And (c) the observables with $\mathcal{D}=0$ correspond to the optimal ones. The first property can be proved by the definition of Euclidean distance in terms of the Frobenius norm and the second is a consequence of Eq. (\ref{error_propagation3}). The last one is just the condition for saturating the low bound given in Eq. (\ref{QCR_ieq2}), which shows that two seemingly different observables might lead to same estimation precision with zero distance between them.
\section{Examples}\label{sec2}
In this section, we apply the theoretical framework into a driven qubit and a bipartite system, where the bipartite system consists of a nucleus and its surrounding electron. In particular, we focus on the estimation of the transition frequency of the qubit and the Larmor frequency of the nucleus. In the first example, we analyze the effect of non-optimal observables on the estimation precision, while for the nucleus and electron, we aim at demonstrating how to optimize the observables of the system with limited accessibility of measurements. In these two examples, the minimum Euclidean distance between the non-optimal observables and the optimal ones are also given.
\subsection{Driven quantum bit}
In this example, we first show the low bound of the estimation of the transition frequency in the driven qubit, then analyze the effect of non-optimal observables on the estimation precision, where the non-optimal observable is simulated by random creation of observables. Finally, we calculate the minimum Euclidean distance between the non-optimal observables and the optimal ones and establish a connection between the distance and the precision of the estimation.

The Hamiltonian of the driven qubit reads,
\begin{eqnarray}
\begin{aligned}
H(\omega_a)=\frac{\omega_a}{2}\sigma_z+F\sigma_x,
\label{Hamiltonian}
\end{aligned}
\end{eqnarray}
where $\omega_a$ is the transition frequency, $\sigma_z$ and $\sigma_x$ are Pauli operators of the qubit and $F$ is the amplitude of the drive. The estimated parameter $\omega_a$ is encoded into an initial density matrix $\rho_0=|\varphi\rangle\langle\varphi|$ by unitary evolution with $U=e^{-iH(\omega_a)t}$
\begin{align}
\rho(\omega_a)=U\rho_0U^\dagger=\frac{1}{2}\left(
                 \begin{matrix}
                   1+\xi &
                   \zeta\\
                   \zeta^*&
                   1-\xi\\
                 \end{matrix}
               \right),
\label{encoded}
\end{align}
where $|\varphi\rangle=\textrm{cos}\phi|g\rangle+\textrm{sin}\phi|e\rangle$, $|g\rangle$ and $|e\rangle$ are the ground and excited states of the qubit, respectively. For simplicity, we have set $\phi=\frac{\pi}{4}$ in Eq. (\ref{encoded}), $\xi=(1-\textrm{cos}\lambda t)\textrm{cos}\vartheta \textrm{sin}\vartheta$, $\zeta=\textrm{cos}^2\vartheta+\textrm{sin}\vartheta(\textrm{sin}\vartheta \textrm{cos}\lambda t-i\textrm{sin}\lambda t)$, $\lambda=\sqrt{4F^2+\omega_a^2}$ and $\vartheta=\textrm{arctan}(\frac{\omega_a}{2F})$.

\begin{figure}[t]
	\centering
	\includegraphics[width=0.525\textwidth]{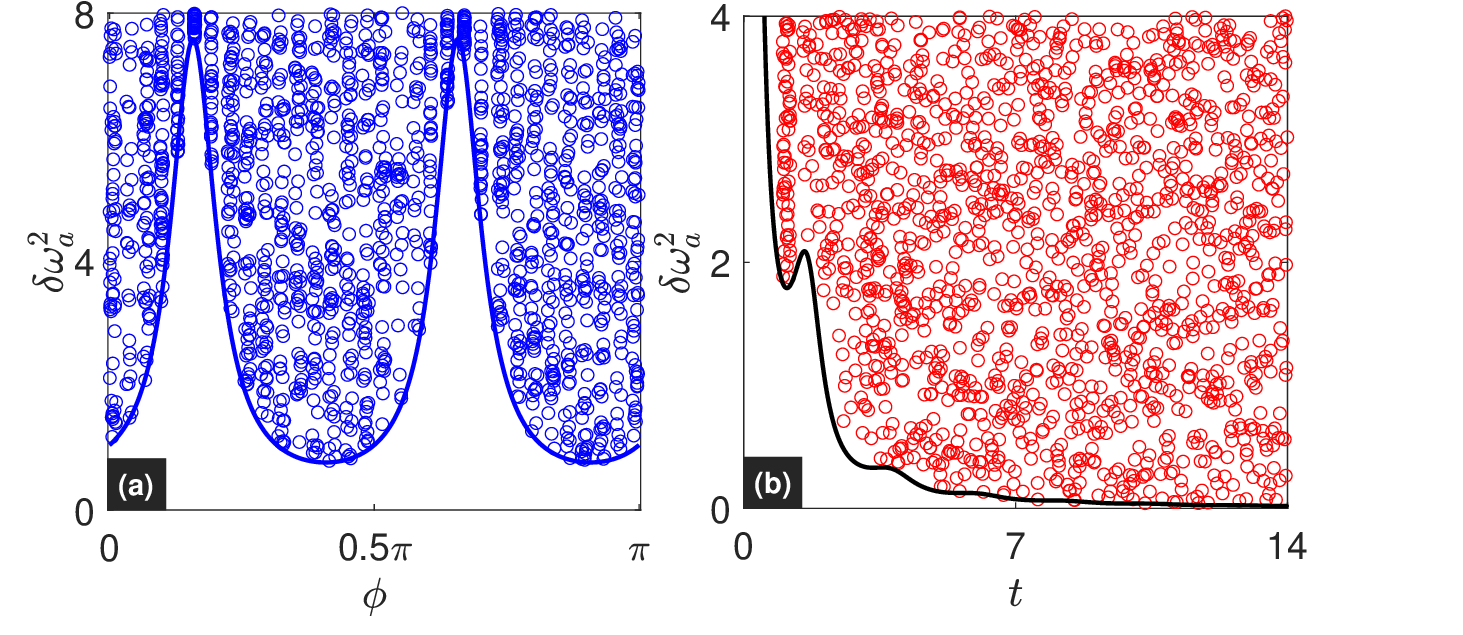}
	\caption {(a) The variance of the estimated transition frequency $\omega_a$ versus $\phi$ for different initial state $|\varphi\rangle=\textrm{cos}\phi|g\rangle+\textrm{sin}\phi|e\rangle$ of the driven qubit. The blue solid line is the low bound of the variance at $t=1.5$.  (b) The variance of $\omega_a$ versus time $t$ in the driven qubit. The black solid line is the low bound with $\phi=\frac{\pi}{4}$. The blue and red circles in (a) and (b) are the results for non-optimal observables $A$, which are chosen randomly via random numbers $\delta\tilde{A}_j$, $j\in\{s,x,y,z\}$. The other system parameters are chosen as $\omega_a=2$ and $F=1$.}
	\label{Driven_bit_mix}
\end{figure}
We calculate the low bound of the variance in Eq. (\ref{QCR_ieq2}) for transition frequency $\omega_a$ and the numerical results are shown in Figs. \ref{Driven_bit_mix}(a) and (b) for different initial states with respect to $\phi$ and time $t$ using blue and black solid lines, respectively. Then we analyze the effect of the non-optimal observables on the estimation precision. For an arbitrary Hermitian non-optimal observable $A=A_{opt}+\delta A$ of the driven qubit, it always can be expressed as
\begin{eqnarray}
\begin{aligned}
A=A_s\mathbb{I}+\sum_iA_i\sigma_i, i\in\{x,y,z\},
\label{A}
\end{aligned}
\end{eqnarray}
where $A_j=\tilde{A}_j+\delta\tilde{A}_j$, $j\in\{s,x,y,z\}$, are the coefficients of the identity operator $\mathbb{I}$ and Pauli operators $\sigma_i$ of the qubit, $\tilde{A}_j$ and $\delta \tilde{A}_j$ correspond to the optimal observable $A_{opt}$ and the deviation $\delta A$, respectively, satisfying $A_{opt}=\tilde{A}_s\mathbb{I}+\sum_i\tilde{A}_i\sigma_i$ and $\delta A=\delta \tilde{A}_s\mathbb{I}+\sum_i\delta\tilde{A}_i\sigma_i$. Next, in order to obtain the variance of $\omega_a$ with non-optimal observable $A$, we give the corresponding quantities in Eq. (\ref{non_optimal0}) with the parameterized density matrix $\rho(\omega_a)$,
\begin{eqnarray}
\begin{aligned}
\langle A\rangle&=A_s+A_x\textrm{Re}(\zeta) -A_y\textrm{Im}(\zeta)+A_z\xi,
\label{meanA1}
\end{aligned}
\end{eqnarray}
\begin{eqnarray}
\begin{aligned}
\langle(\Delta A)^2\rangle=A_x^2+A_y^2+A_z^2-(\gamma_1-\gamma_2)^2,
\label{particalA}
\end{aligned}
\end{eqnarray}
where $\gamma_1=\beta_2\textrm{sin}\vartheta \textrm{cos}\lambda t-\beta_1\textrm{cos}\vartheta$, $\gamma_2=A_y\textrm{sin}\vartheta \textrm{sin}\lambda t$ and $\beta_1=A_x\textrm{cos}\vartheta+A_z\textrm{sin}\vartheta$, $\beta_2=A_z\textrm{cos}\vartheta-A_x\textrm{sin}\vartheta$. The first
\begin{figure}[t]
\centering
\includegraphics[width=0.525\textwidth]{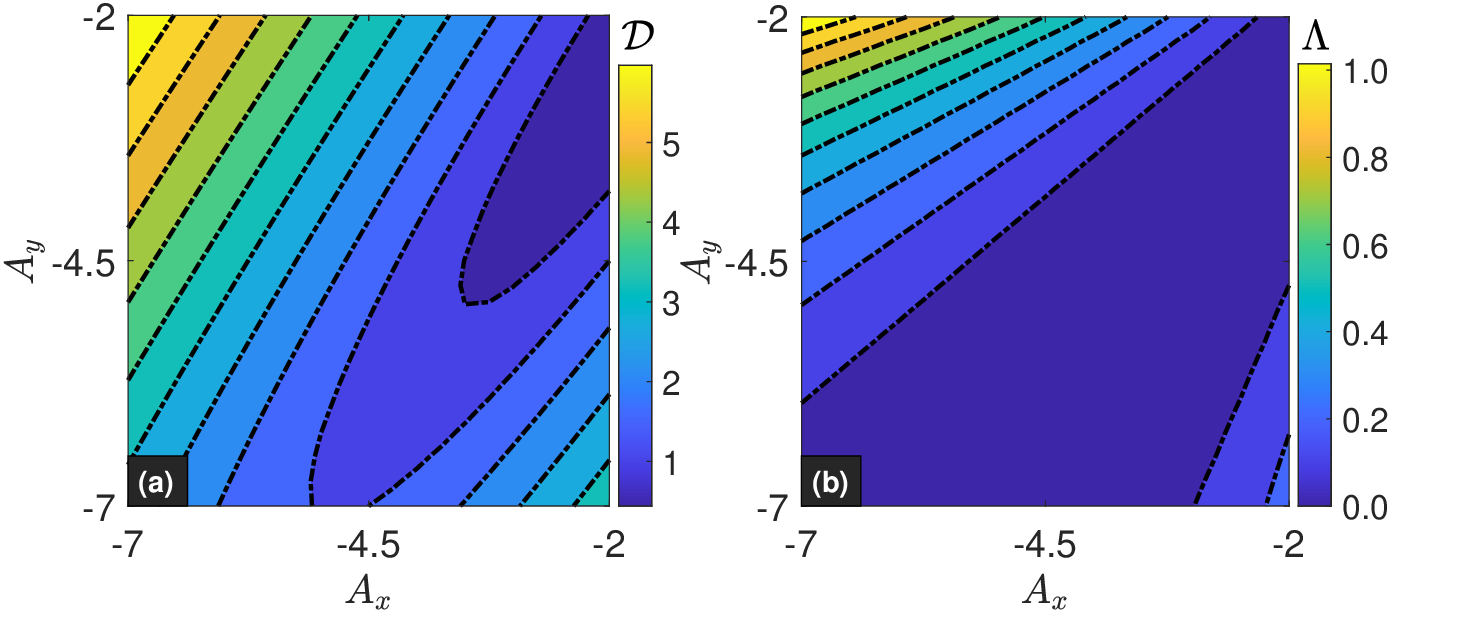}
    \caption{(a) Contour plots for the minimum Euclidean distance $\mathcal{D}$ of non-optimal observables $A$ from the optimal ones $\mathcal{L}_{\omega_a}$. (b) Contour plots for the quantity $\Lambda$ in Eq. (\ref{error_propagation3}) with observables $A$. The parameters chosen are $A_s=-0.7, A_z=0.2, t=1$ and the others are the same as in Fig. \ref{Driven_bit_mix}.}
    \label{two_com}
\end{figure}
derivative of $\langle A\rangle$ with respect to $\omega_a$ is
\begin{eqnarray}
\begin{aligned}
\frac{\partial\langle A\rangle}{\partial\omega_a}=\frac{\partial\textrm{Re}(\zeta)}{\partial\omega_a}A_x-\frac{\partial\textrm{Im}(\zeta)}{\partial\omega_a}A_y+\frac{\partial\xi}{\partial\omega_a}A_z,
\label{particalA}
\end{aligned}
\end{eqnarray}
where
\begin{eqnarray}
\begin{aligned}
\frac{\partial\textrm{Re}(\zeta)}{\partial\omega_a}=-\frac{\textrm{cos}\vartheta\xi}{\lambda}-\frac{\textrm{sin}^3\vartheta\textrm{sin}\lambda t}{2}t,\nonumber
\label{where1}
\end{aligned}
\end{eqnarray}
\begin{eqnarray}
\begin{aligned}
\frac{\partial\textrm{Im}(\zeta)}{\partial\omega_a}=-\frac{\textrm{cos}^2\vartheta \textrm{sin}\lambda t}{2\lambda}-\frac{t\textrm{sin}^2\vartheta \textrm{cos}\lambda}{2}t,\nonumber
\label{where2}
\end{aligned}
\end{eqnarray}
\begin{eqnarray}
\begin{aligned}
\frac{\partial\xi}{\partial\omega_a}=\frac{\xi \textrm{cos}2\vartheta}{2\lambda\textrm{sin}\vartheta}+\frac{t\textrm{sin}^2\vartheta\textrm{cos}\vartheta\textrm{sin}\lambda}{2}t.\nonumber
\label{where3}
\end{aligned}
\end{eqnarray}
We generate random numbers for $\delta\tilde{A}_j$ to simulate the non-optimal observables in realistic quantum measurements. And the variance of $\omega_a$ for the non-optimal observables calculated with the error propagation formula are shown in Figs. \ref{Driven_bit_mix}(a) and (b) with blue and red circles, respectively. As expected, the precision obtain by the non-optimal observables is no better than the optimal ones which is limited by $\Lambda\geq0$.

Finally, we analyze the minimum Euclidean distance in Eq. (\ref{distance}) between the non-optimal observables and the optimal ones $\mathcal{L}_{\omega_a}$, where $\mathcal{L}_{\omega_a}$ are commutative with SLD $L_{\omega_a}$ satisfying $\frac{\partial\rho(\omega_a)}{\partial\omega_a}=\frac{1}{2}[\rho(\omega_a) L_{\omega_a}+L_{\omega_a}\rho(\omega_a)]$, and the results are shown in Fig. \ref{two_com}(a) (the generation of $\mathcal{L}_{\omega_a}$ is shown in Appendix \ref{appendixC}). For comparison, we also give the result of $\Lambda$ [the validity of $\Lambda$ in Eq. (\ref{error_propagation3}) is shown in Appendix \ref{appendixA1}] in Fig. \ref{two_com}(b) which shows that the observable with a shorter minimum Euclidean distance to the optimal ones benefits the estimation precision. This provides us with a criterion to find an suitable observable for the estimation.

\subsection{Bipartite system}
In the last subsection, we have analyzed the effect of non-optimal observables on the estimation precision in a driven qubit. In this subsection, we will present a scheme to optimize the observables in a coupled nucleus-electron system to estimate the nuclear Larmor frequency when the optimal global measurement is inaccessible. This problem comes from the fact that directly measuring the nuclear Larmor frequency in coupled nucleus-electron system \cite{Jelezko2006,Lee2013,Shi2014} is difficult and the control of nucleus in solids have attracted a great deal of interest, such as in silicon \cite{Pla2013,Morton2008}, silicon carbide \cite{Ivady2015,Falk2015} and NV-centers in diamond \cite{Unden2016,Pfender2019,Whaites2022}. In the following, we will consider the NV-center in diamond as the bipartite system, where the NV-center electronic spin is coupled to a single nucleus via the dipole-dipole interaction for simplicity, rather than a pair of (or more) NV centers, and we will focus on the estimation of the nuclear Larmor frequency via measurements on the electronic observables. Indeed, we achieve a precise estimation by optimizing only the observables of the electronic spin, which corresponds to local measurement in the subsystem. Furthermore, we also discuss the optimization of the joint observables in the bipartite system. Finally, we show the low bounds of the estimation variance by performing the local measurements in different subsystems and compare that with the QCRB of the whole system.

In order to demonstrate the optimization of the observable in this bipartite system concisely simplify both
analytical and numerical calculations, we consider the nucleus and electron as spin $S=\frac{1}{2}$ systems. And the Hamiltonian of the coupled nucleus-electron system via the dipole-dipole interaction under an on-resonance drive is given by \cite{Tannoudji1986,Aharon2019}
\begin{eqnarray}
\begin{aligned}
H=\frac{\omega_0}{2}\sigma_z+\frac{\omega_l}{2}I_z+g\sigma_zI_x+\Omega_1\sigma_x\textrm{cos}(\omega_0t),
\label{NV_H}
\end{aligned}
\end{eqnarray}
where $\omega_0$ and $\omega_l$ are the energy level spacing of the electron spin and the Larmor frequency of the nucleus, respectively. $\sigma_{x,z}$ and $I_{x,z}$ denote respectively the Pauli operators of the electron and the nucleus. $\Omega_1$ is the Rabi frequency of the drive and $g$ is the nucleus-electron coupling strength. We move to the interaction picture with respect to the first term, $H_0=\frac{\omega_0}{2}\sigma_z$, to eliminate the time-dependence of the Hamiltonian. By the rotating-wave-approximation with the assumption that $\omega_0\gg\Omega_1$ \cite{Aharon2019,Allen1987}, the effective Hamiltonian reads
\begin{eqnarray}
\begin{aligned}
H_I(\omega_l)=\frac{\Omega_1}{2}\sigma_x+\frac{\omega_l}{2}I_z+g\sigma_zI_x.
\label{RWA_NV_H}
\end{aligned}
\end{eqnarray}
In the following, we consider the unitary dynamic with Hamiltonian (\ref{RWA_NV_H}) to encode the estimated parameter $\omega_l$ into an initial density matrix $\tilde{\rho}_0=|\varphi^n\rangle\langle\varphi^n|\otimes|\varphi^e\rangle\langle\varphi^e|$, here $|\varphi^n\rangle=\textrm{cos}\phi_1|g^n\rangle+\textrm{sin}\phi_1|e^n\rangle$ and $|\varphi^e\rangle=\textrm{cos}\phi_2|g^e\rangle+\textrm{sin}\phi_2|e^e\rangle$ are the initial states for nucleus and electron, respectively. $|g^i\rangle$ and $|e^i\rangle$, $i\in\{e,n\}$, are the ground and excited states of electronic and nuclear spin. The parameterized density matrix is
\begin{eqnarray}
\begin{aligned}
\tilde{\rho}(\omega_l)=\tilde{U}\tilde{\rho}_0\tilde{U}^\dagger=|\Phi(\omega_l)\rangle\langle\Phi(\omega_l)|,
\label{encoded2}
\end{aligned}
\end{eqnarray}
where $\tilde{U}=e^{-iH_I(\omega_l)t}$ is the unitary operator, and
\begin{widetext}
\begin{eqnarray}
\begin{aligned}
|\Phi(\omega_l)\rangle=\frac{1}{\sqrt{2}}\left(e^{-iE_1t}\textrm{cos}\theta_+|E_1\rangle+e^{-iE_2t}\textrm{sin}\varphi_+|E_2\rangle+e^{-iE_3t}\textrm{cos}\theta_-|E_3\rangle+e^{-iE_4t}\textrm{sin}\varphi_-|E_4\rangle\right),
\label{Phi}
\end{aligned}
\end{eqnarray}
\end{widetext}
where $\textrm{tan}\theta_+=\frac{g}{\alpha+\sqrt{\alpha^2+g^2}}$, $\textrm{tan}\varphi_+=\frac{g}{\beta+\sqrt{\beta^2+g^2}}$, $\textrm{tan}\theta_-=\frac{g}{\alpha-\sqrt{\alpha^2+g^2}}$ and $\textrm{tan}\varphi_-=\frac{g}{\beta-\sqrt{\beta^2+g^2}}$.  $\alpha=\frac{\omega_l-\Omega_1}{2}$ and $\beta=\frac{\omega_l+\Omega_1}{2}$, and we have set $\phi_1=\phi_2=\frac{\pi}{4}$ in Eq. (\ref{Phi}) for simplicity. $E_i$ and $|E_i\rangle$ are the eigenvalues and eigenvectors of the Hamiltonian (\ref{RWA_NV_H}), respectively. The detail calculation of Eq. (\ref{Phi}) can be found in Appendix \ref{appendixD}.

\begin{figure}[t]
\centering
\includegraphics[width=0.51\textwidth]{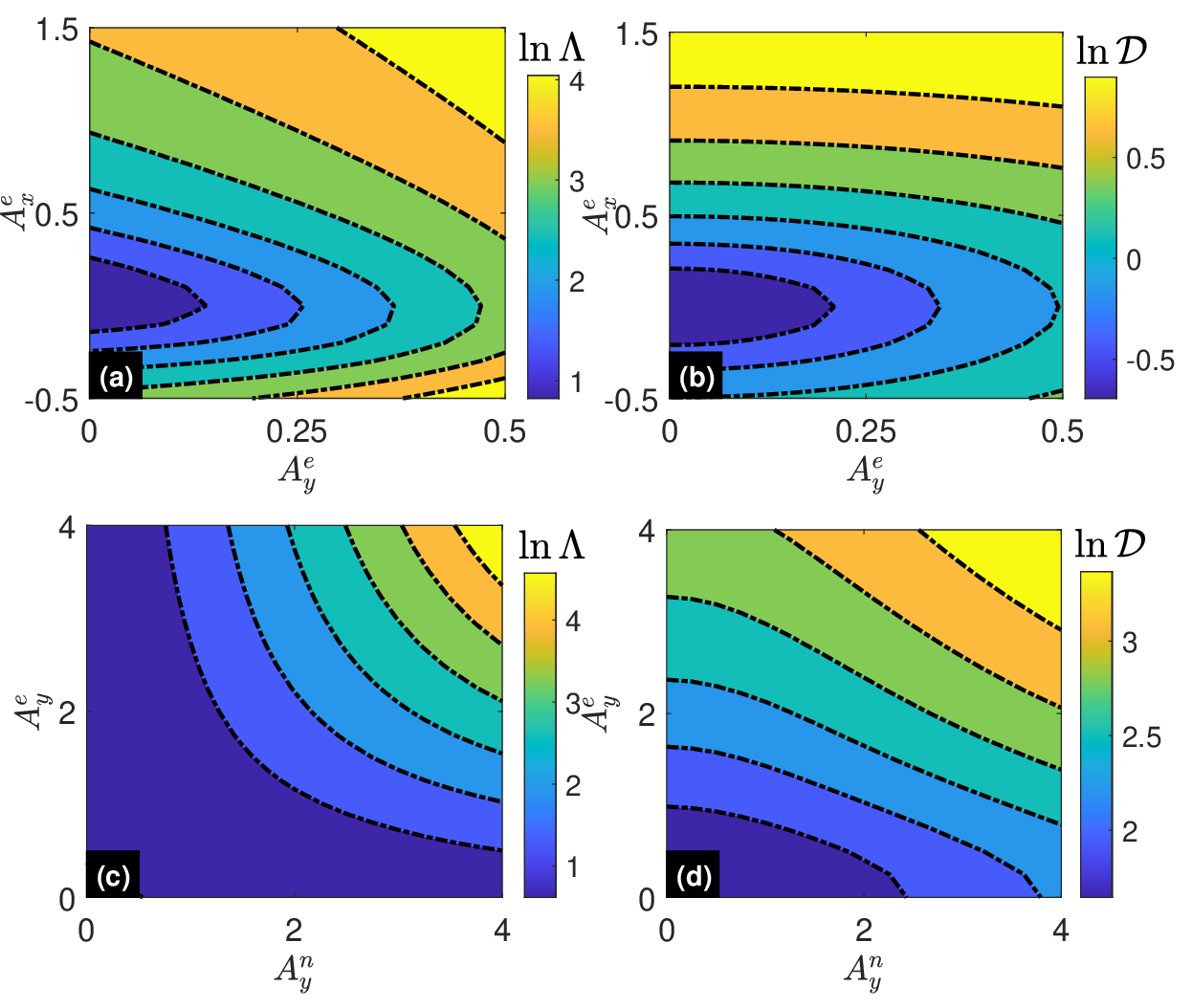}
    \caption{(a) Contour plots for $\textrm{ln}\Lambda$ in Eq. (\ref{error_propagation3}) for different local observables $\mathcal{A}^e$. (b) Contour plots for the minimum Euclidean distance $\textrm{ln}\mathcal{D}$ between the local observable $\mathcal{A}^e$ and the optimal ones $\mathcal{L}_{\omega_l}$ in coupled nucleus-electron system. The parameters are $A^e_s=-1, A^e_z=-0.25, t=2$ for panels (a) and (b). (c) $\textrm{ln}\Lambda$ for different joint observables $\mathcal{A}^{n\otimes e}$. (d) The minimum Euclidean distance $\textrm{ln}\mathcal{D}$ between the joint observables $\mathcal{A}^{n\otimes e}$ and $\mathcal{L}_{\omega_l}$ for coupled nucleus-electron system. The parameters are $A^n_s=2$, $A^n_x=1$, $A^n_z=-1$, $A^e_s=0$, $A^e_x=1$, $A^e_z=-0.5$ and $t=2$ for panels (c) and (d). Other parameters are chosen as $\phi_1=\phi_2=\frac{\pi}{4}$, $\Omega_1=3$, $\omega_l=2$, and $g=2$.}
\label{four_com}
\end{figure}

Next, we consider the optimization of the local observables in electron subsystem to achieve a precise estimation for the nuclear Larmor frequency. An arbitrary observable of the electron subsystem can be expressed as
\begin{eqnarray}
\begin{aligned}
\mathcal{A}^e=\mathbb{I}^n\otimes A^e,
\label{Ae}
\end{aligned}
\end{eqnarray}
where $A^e=A^e_s\mathbb{I}^e+A^e_x\sigma_x+A^e_y\sigma_y+A^e_z\sigma_z$, and $A^e_i$, $i\in\{s,x,y,z\}$ are the coefficients of identity operator $\mathbb{I}^e$ and Pauli operators $\sigma_i$ of the electron spin. We optimize the local observables and show the result of $\textrm{ln} \Lambda$ in Fig. \ref{four_com}(a). In addition, we calculate the minimum Euclidean distance between the local observable $\mathcal{A}^e$ and the optimal ones $\mathcal{L}_{\omega_l}$ (commute with $L_{\omega_l}$ satisfying $\frac{\partial\tilde{\rho}(\omega_l)}{\partial\omega_l}=\frac{1}{2}[\tilde{\rho}(\omega_l) L_{\omega_l}+L_{\omega_l}\tilde{\rho}(\omega_l)]$) and show the results in Fig. \ref{four_com}(b). We find that the observables with shorter minimum Euclidean distance to the optimal ones benefits the estimation precision. And the discussions of the effect of noise on the estimation precision is shown in the Appendix \ref{appendixE}.

Furthermore, we discuss the optimization of the joint observables
\begin{eqnarray}
\begin{aligned}
\mathcal{A}^{n\otimes e}=A^n\otimes A^e,
\label{composite}
\end{aligned}
\end{eqnarray}
in both electron and nucleus subsystems, where $A^n=A^n_s\mathbb{I}^n+A^n_xI_x+A^n_yI_y+A^n_zI_z$, and $A^n_j$, $j\in\{s,x,y,z\}$, are the coefficients of identity operator $\mathbb{I}^n$ and Pauli operators $I_i$ of nucleus. We show $\textrm{ln}\Lambda$ of the joint observable in Fig. \ref{four_com}(c) and the minimum Euclidean distance between the joint observables and the optimal ones in Fig. \ref{four_com}(d). It is clearly that the observable with shorter minimum Euclidean distance to the optimal ones benefits the estimation precision.

\begin{figure}[t]
	\centering
	\includegraphics[width=0.525\textwidth]{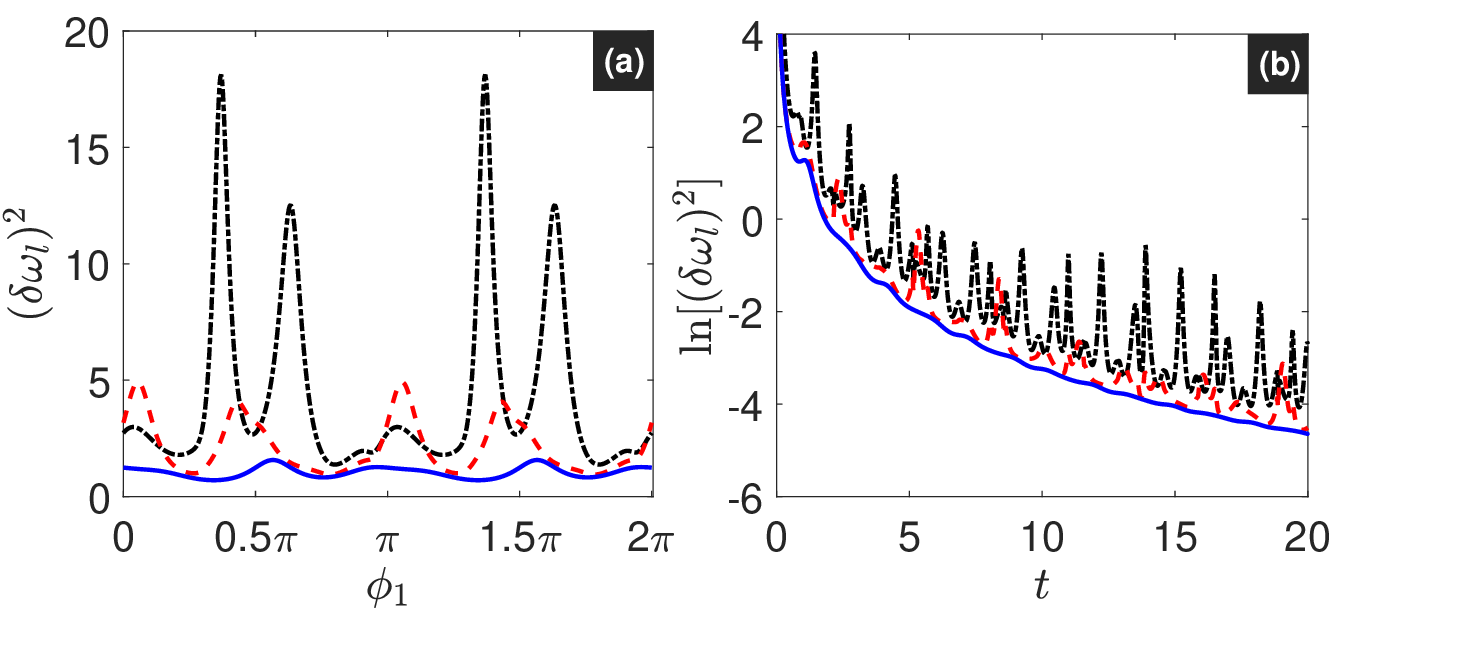}
	\caption {(a) and (b) show the variance of the Larmor frequency $\omega_l$ versus $\phi_1$ (we have set $\phi_1=\phi_2$) and time $t$, respectively. In (a) and (b), the blue solid lines are the low bounds of the variance in the global system. The red dashed lines and black dashed-dot lines are the bounds for the nucleus and electron subsystems, respectively. Other parameters are the same as in Fig. \ref{four_com}.}
	\label{Fin_nul}
\end{figure}

Finally, we calculate the low bound of the variance of $\omega_l$ for the global system and the numerical results are shown for different initial states with respect to $\phi_1$ and time $t$ using blue solid lines in Figs. \ref{Fin_nul}(a) and (b), where we have set $\phi_1=\phi_2$ in the initial density matrix $\tilde{\rho}_0$ for simplicity. In addition, we also give the low bounds of the variance for the local observables in nucleus and electron subsystems with red dashed lines and black dashed-dot lines, respectively, which is the inverse of the subsystem QFI
\begin{eqnarray}
\begin{aligned}
\mathcal{F}^i_Q=\textrm{Tr}[\rho^i_{\omega_l} (L^i_{\omega_l})^2],
\label{sub_QFI}
\end{aligned}
\end{eqnarray}
where $\rho_{\omega_l}^i=\textrm{Tr}_{\neq i}[\tilde{\rho}(\omega_l)]$, $i\in\{e, n\}$, are the reduced density matrix of electronic and nuclear subsystem, respectively. $L^i_{\omega_l}$ are the SLDs for each subsystem satisfying $\frac{\partial\rho^i_{\omega_l}}{\partial\omega_l}=\frac{1}{2}(\rho^i_{\omega_l} L^i_{\omega_l}+L^i_{\omega_l}\rho^i_{\omega_l})$ (see Appendix \ref{appendixB} for the details). It is clear that the precision obtained by the partial accessibility is no better than that by whole accessibility of the measurements. And, we point out there is no specific relation for which subsystem observable is better for the estimation. Therefore, one should carefully choose an suitable subsystem and observable to obtain a precise estimation if all the measurements in the subsystem are accessible.

\section{Conclusion}\label{sec3}
In this paper, we presented a theoretical framework to analyze the effect of the non-optimal observables on the precision of parameter estimation, and proposed a scheme to optimize the observables with limited accessibility of measurements. To be specific, in order to obtain a higher precision in composite systems with all measurements restricted in the subsystems, we have defined a quantity $\Lambda$ to characterize the effect of the non-optimal measurements on the estimation precision and proved it is always non-negative. In addition, we have showed in details how to optimize the local and joint observables. To establish the connection between the non-optimal measurements and precision of the parameter estimation, we have introduced the minimum Euclidean distance to characterize the non-optimality of the measurements, and found that the measurement with a shorter minimum Euclidean distance to the optimal ones benefits the estimation precision.

As an example, we applied the theory to analyze the effect of non-optimal observables on the estimation precision in a driven qubit and a bipartite system. In the bipartite system consisting of an electron and a nucleus, we try to estimate the nuclear Larmor frequency via the measurements of the electron degree of freedom. The results suggested that the precision obtained by the partial accessibility of the measurements is no better than that by whole accessibility. In those two examples, we also analyzed a minimum Euclidean distance between the observables and the optimal ones to characterize the quantity of the non-optimal measurement. We show that the measurement with shorter minimum Euclidean distance to the optimal ones benefits the estimation precision, which provides us with a criterion to find an suitable measurement for the estimation in case of the optimal ones are inaccessible.

\section{acknowledgments}\label{sec5}
This work is supported by National Natural Science Foundation of China (NSFC) under Grants No. 12175033, No. 12147206 and National Key R$\&$D Program of China No. 2021YFE0193500.

\begin{appendix}
\section{the sufficient condition of observable $A$ for saturating QCRB}\label{appendixA}
In this appendix, we give the sufficient condition of the optimal observable $A_{opt}$ for saturating the QCRB briefly. According to error propagation formula (\ref{non_optimal0}) and the Schr\"{o}dinger-Robertson uncertainty relation for two arbitrary Hermitian operators $X$ and $Y$ \cite{Robertson1929,Robertson1934},
\begin{eqnarray}
\begin{aligned}
\langle(\Delta X)^2\rangle\langle(\Delta Y)^2\rangle\geq \textrm{Cov}^2(X,Y)+\frac{|\langle[X,Y]\rangle|^2}{4},
\label{SHUR}
\end{aligned}
\end{eqnarray}
where $\textrm{Cov}(X,Y)=\frac{\langle\{X,Y\}\rangle}{2}-\langle X\rangle\langle Y\rangle$ is the covariance of the Hermitian operators, $[X,Y]=XY-YX$ and $\{X,Y\}=XY+YX$ are the commutator and anti-commutator, respectively. One can obtain the following inequality
\begin{eqnarray}
\begin{aligned}
(\delta\theta)^2&=\frac{\langle(\Delta A)^2\rangle\langle(\Delta L_\theta)^2\rangle}{|\partial_\theta\langle A\rangle|^2\langle(\Delta L_\theta)^2\rangle}\\&\geq\frac{\textrm{Cov}^2(A,L_\theta)}{|\partial_\theta\langle A\rangle|^2\langle(\Delta L_\theta)^2\rangle}+\frac{|\langle[A,L_\theta]\rangle|^2}{4|\partial_\theta\langle A\rangle|^2\langle(\Delta L_\theta)^2\rangle}\\&=\frac{1}{\mathcal{F}_Q}\left(1+\frac{|\langle[A,L_\theta]\rangle|^2}{\langle\{A,L_\theta\}\rangle^2}\right)\\&\geq\frac{1}{\mathcal{F}_Q},
\label{QCR_ieq}
\end{aligned}
\end{eqnarray}
where $L_\theta$ is the SLD satisfies $\frac{\partial\rho_\theta}{\partial\theta}=\frac{1}{2}(\rho_\theta L_\theta+L_\theta\rho_\theta)$, which implies $\langle L_\theta\rangle=0$. And $\mathcal{F}_Q=\textrm{Tr}(\rho_\theta L_\theta^2)$=$\langle(\Delta L_\theta)^2\rangle$ is the QFI. A sufficient condition for taking the low bound in Eq. (\ref{QCR_ieq}) is the optimal $A_{opt}$ satisfying $[A_{opt},L_\theta]=0$.

Next, we prove that the quantity $\Lambda=(\delta\theta)^2-\textrm{min}[(\delta\hat{\theta})^2]$ defined in Eq. (\ref{error_propagation3}) is always non-negative $\Lambda\geq0$. For a non-optimal observable $A=A_{opt}+\delta A$, Eq. (\ref{QCR_ieq}) can be re-written as
\begin{eqnarray}
\begin{aligned}
(\delta\theta)^2\geq\frac{1}{\mathcal{F}_Q}\left(1+\frac{|\langle[\delta A,L_\theta]\rangle|^2}{\langle\{A,L_\theta\}\rangle^2}\right),
\label{QCR_ieq3}
\end{aligned}
\end{eqnarray}
where $[A_{opt},L_\theta]=0$ has been used. Therefore,
\begin{eqnarray}
\begin{aligned}
\Lambda\geq\frac{|\langle[\delta A,L_\theta]\rangle|^2}{\mathcal{F}_Q\langle\{A,L_\theta\}\rangle^2}\geq0,
\label{QCR_ieq3}
\end{aligned}
\end{eqnarray}
where we have shown the non-negativity of $\Lambda$.
\section{QFI for the subsystems}\label{appendixB}
The total Hamiltonian of a composite system can be expressed as Eq. (\ref{general_H}). And the unknown parameter $\theta$ is encoded into the initial density matrix $\rho_0$ by $\rho_\theta=U\rho_0U^\dagger$, where $U=e^{-iH(\theta)t}$ is the unitary operator. Therefore, one can obtain the reduced density matrix as
\begin{eqnarray}
\begin{aligned}
\rho^{i}_\theta=\textrm{Tr}_{\neq i}(\rho_\theta).
\label{reduce}
\end{aligned}
\end{eqnarray}
By defining the SLD operator $L^i_\theta$, $i\in\{a,b\}$, which obeys the operator equation \cite{Braunstein1994,Braunstein1996}
\begin{eqnarray}
\begin{aligned}
\frac{\partial\rho^i_\theta}{\partial\theta}=\frac{1}{2}(\rho^i_\theta L^i_\theta+L^i_\theta\rho^i_\theta),
\label{SLD_def_sub}
\end{aligned}
\end{eqnarray}
one can obtain the expression of the SLD with the spectral decomposition of the density matrix $\rho^i_\theta=\sum_n\mu^i_n|\psi^i_n\rangle\langle\psi^i_n|$, which can be expressed as
\begin{eqnarray}
\begin{aligned}
L^i_\theta=2\sum_{n,m}\frac{\langle\psi^i_n|\partial_\theta\rho^i_\theta|\psi^i_m\rangle}{\mu^i_n+\mu^i_m}|\psi^i_n\rangle\langle\psi^i_m|,
\label{SLDC3}
\end{aligned}
\end{eqnarray}
where $\mu^i_n$ and $|\psi^i_n\rangle$ are the eigenvalues and eigenvectors of $\rho^i_\theta$ and  $\mu^i_n+\mu^i_m\neq0$. The subsystem QFI is
\begin{eqnarray}
\begin{aligned}
\mathcal{F}^i_Q=\textrm{Tr}[\rho^i_\theta (L^i_\theta)^2],
\label{QFI_sub}
\end{aligned}
\end{eqnarray}
for different subsystems.
\section{Construct an arbitrary observable commutating with the known one}\label{appendixC}
In this appendix, we prove that it is necessary to employ two parameters to describe an arbitrary Hermitian operator commutating with another one for a qubit. This proof provides a reference for the generation of an arbitrary observable commutating with another known one.

For a given Hermitian operator $G$, the matrix form is
\begin{eqnarray}
\begin{aligned}
G\!=\!\left(
                 \begin{array}{cc}
                   a & b+ci \\
                 b-ci & d \\
                 \end{array}
               \right),
\label{A1}
\end{aligned}
\end{eqnarray}
where $a, b, c, d$ are real known parameters. An arbitrary unknown Hermitian operator $K$ commutating with $G$ is denoted as
\begin{eqnarray}
\begin{aligned}
K\!=\!\left(
                 \begin{array}{cc}
                   p & r+si \\
                 r-si & q \\
                 \end{array}
               \right),
\label{A2}
\end{aligned}
\end{eqnarray}
where $p, r, s, q$ are real undetermined parameters. According to $[G,K]=0$, we could obtain two independent equations as follows
\begin{eqnarray}
\begin{aligned}
(q-p)c&=(d-a)s,\\
(q-p)b&=(d-a)r.
\label{A3}
\end{aligned}
\end{eqnarray}
The two equations constraint the undetermined parameters to two which means there are two unknown parameters for an arbitrary Hermitian operator commutating with another one. Therefore, we construct an arbitrary operator commutating with $G$ as following,
\begin{eqnarray}
\begin{aligned}
K=\sum_{i=1,2}k_i|\phi_i\rangle\langle\phi_i|,
\label{A4}
\end{aligned}
\end{eqnarray}
where $k_i$ are the arbitrary real parameters and $|\phi_i\rangle$ is the eigenvectors of $G$.
\begin{widetext}
\section{the validity of Eq. (\ref{error_propagation3})}\label{appendixA1}
To check the validity of  Eq. (\ref{error_propagation3}), we show the numerical results for the relevant quantities in Fig. \ref{reply_1} for the driven qubit in the first example, where the blue dashed line is the result of $\Lambda$ and the orange dashed-dot line is $\delta \omega_a^2$ in Eq. (\ref{non_optimal0}) for different non-optimal measurements. The difference between $\delta \omega_a^2$ and $\Lambda$ is shown with yellow solid line, which is in good agreement with the results of the low bound of the precision $\frac{1}{\mathcal{F}_Q}$. This means Eq. (\ref{error_propagation3}) is good  to quantify the effect of the non-optimal measurements on the best estimation precision.
\begin{figure}[h]
	\centering
	\includegraphics[width=0.45\textwidth]{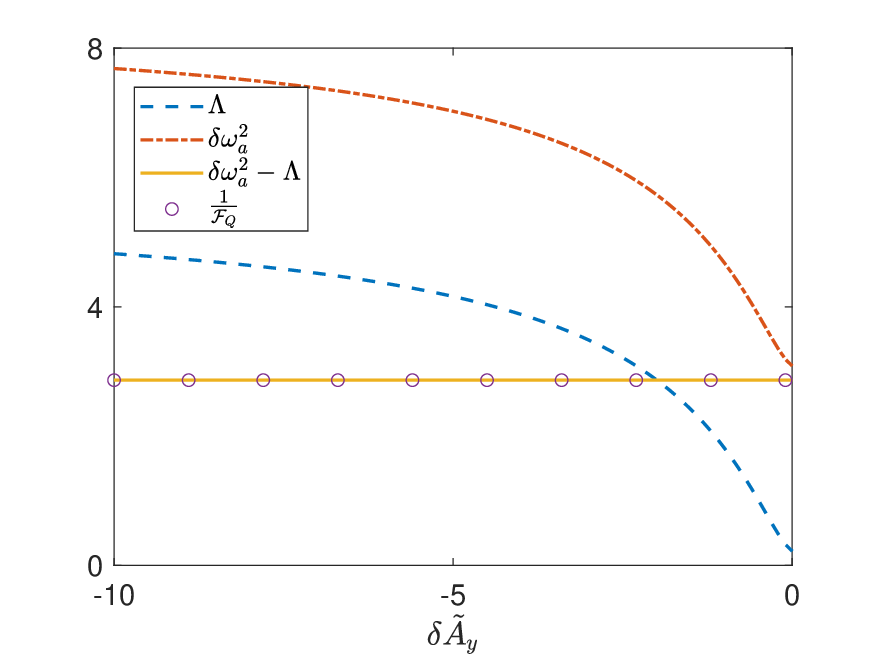}
	\caption {$\delta \omega_a^2$ and $\Lambda$ are calculated by Eq. (\ref{non_optimal0}) and Eq. (\ref{error_propagation3}), respectively, for non-optimal observable $A=A_{opt}+\delta A$, where $\delta A=\delta \tilde{A}_s\mathbb{I}+\sum_i\delta\tilde{A}_i\sigma_i$, $i\in\{x,y,z\}$, $\mathbb{I}$ is the identity operator and $\sigma_i$ are the Pauli operators of the driven qubit. The purple circles are the results of $\frac{1}{\mathcal{F}_Q}$, where $\mathcal{F}_Q$ is the quantum Fisher information $\mathcal{F}_Q=\textrm{Tr}(\rho_{\omega_a} L_{\omega_a}^2)$. The results show that the difference between $\delta \omega_a^2$ and $\Lambda$ is according well with the low bound of the precision $\frac{1}{\mathcal{F}_Q}$. The system parameters chosen are $t=1$ and others are the same as in Fig. \ref{Driven_bit_mix}.}
	\label{reply_1}
\end{figure}
\section{solve the eigenvalues and eigenvectors for Hamiltonian (\ref{RWA_NV_H})}\label{appendixD}
Here, we give the details expression of $|\Phi(\omega_l)\rangle$ in Eq. (\ref{Phi}) by solving the eigenvectors of the Hamiltonian in Eq. (\ref{RWA_NV_H}). We can obtain the matrix form of the Hamiltonian (\ref{RWA_NV_H}) in the basis $|e^e,g^n\rangle$, $|e^e,e^n\rangle$, $|g^e,g^n\rangle$ and $|g^e,e^n\rangle$,
\begin{eqnarray}
\begin{aligned}
H_I\!=\left(
                 \begin{array}{cccc}
                   -A & g & B & 0 \\
                    g & A & 0 & B \\
                    B & 0 &-A &-g \\
                    0 & B &-g & A \\
                 \end{array}
               \right),
\label{nul_H}
\end{aligned}
\end{eqnarray}
where $|g^i\rangle$ and $|e^i\rangle$, $i\in\{e,n\}$, are the ground and excited states of electronic and nuclear spin, respectively, and $A=\frac{\omega_l}{2}$, $B=\frac{\Omega_1}{2}$. The normalized eigenvectors are
\begin{eqnarray}
\begin{aligned}
|E_1\rangle&=\frac{\sqrt{2}}{2}\left(
                                \begin{array}{c}
                                  \textrm{cos}\theta_+ \\
                                  -\textrm{sin}\theta_+ \\
                                  \textrm{cos}\theta_+ \\
                                  \textrm{sin}\theta_+ \\
                                \end{array}
                              \right),
|E_2\rangle=\frac{\sqrt{2}}{2}\left(
                                \begin{array}{c}
                                  -\textrm{cos}\varphi_+ \\
                                  \textrm{sin}\varphi_+ \\
                                  \textrm{cos}\varphi_+ \\
                                  \textrm{sin}\varphi_+ \\
                                \end{array}
                              \right),\\
|E_3\rangle&=\frac{\sqrt{2}}{2}\left(
                                \begin{array}{c}
                                  \textrm{cos}\theta_- \\
                                  -\textrm{sin}\theta_- \\
                                  \textrm{cos}\theta_- \\
                                  \textrm{sin}\theta_- \\
                                \end{array}
                              \right),
|E_4\rangle=\frac{\sqrt{2}}{2}\left(
                                \begin{array}{c}
                                  -\textrm{cos}\varphi_- \\
                                  \textrm{sin}\varphi_- \\
                                  \textrm{cos}\varphi_- \\
                                  \textrm{sin}\varphi_- \\
                                \end{array}
                              \right),
\label{eigenstates}
\end{aligned}
\end{eqnarray}
where
\begin{eqnarray}
\begin{aligned}
\textrm{tan}\theta_+=\frac{g}{\alpha+\sqrt{\alpha^2+g^2}},\textrm{tan}\varphi_+=\frac{g}{\beta+\sqrt{\beta^2+g^2}},\\ \textrm{tan}\theta_-=\frac{g}{\alpha-\sqrt{\alpha^2+g^2}},\textrm{tan}\varphi_-=\frac{g}{\beta-\sqrt{\beta^2+g^2}},
\label{four}
\end{aligned}
\end{eqnarray}
$\alpha=A-B$ and $\beta=A+B$. In addition, $\theta_+$, $\varphi_+$, $\theta_-$ and $\varphi_-$ satisfy
\begin{eqnarray}
\begin{aligned}
\textrm{cos}\theta_+ \textrm{cos}\theta_-+\textrm{sin}\theta_+ \textrm{sin}\theta_-=0,\\
\textrm{cos}\varphi_+ \textrm{cos}\varphi_-+\textrm{sin}\varphi_+ \textrm{sin}\varphi_-=0.
\label{relations}
\end{aligned}
\end{eqnarray}
The corresponding eigenvalues are
\begin{eqnarray}
\begin{aligned}
E_1=-\sqrt{\alpha^2+g^2}, E_2=-\sqrt{\beta^2+g^2},\\
E_3=\sqrt{\alpha^2+g^2}, E_4=\sqrt{\beta^2+g^2}.
\label{eigenvalues}
\end{aligned}
\end{eqnarray}
Now, we have given all the eigenvalues and eigenvectors for the Hamiltonian (\ref{RWA_NV_H}). One could obtain the parameterized states by projecting the initial state to the eigenvectors. After some derivations and substitutions, we obtain the parameterized states $|\Phi(\omega_l)\rangle$ in Eq. (\ref{Phi}). Finally, we obtain the following expression
\begin{eqnarray}
\begin{aligned}
|\Phi(\omega_l)\rangle=&\frac{1}{\sqrt{2}}\left(e^{-iE_1t}\textrm{cos}\theta_+|E_1\rangle+e^{-iE_2t}\textrm{sin}\varphi_+|E_2\rangle+e^{-iE_3t}\textrm{cos}\theta_-|E_3\rangle+e^{-iE_4t}\textrm{sin}\varphi_-|E_4\rangle\right)\\ =&\frac{1}{2}\left(
                                \begin{array}{c}
e^{-iE_1t}\textrm{cos}^2\theta_+-e^{-iE_2t}\textrm{sin}\varphi_+\textrm{cos}\varphi_++e^{-iE_3t}\textrm{cos}^2\theta_--e^{-iE_4t}\textrm{sin}\varphi_-\textrm{cos}\varphi_-\\
-e^{-iE_1t}\textrm{sin}\theta_+\textrm{cos}\theta_++e^{-iE_2t}\textrm{sin}^2\varphi_+-e^{-iE_3t}\textrm{sin}\theta_-\textrm{cos}\theta_-+e^{-iE_4t}\textrm{sin}^2\varphi_-\\
e^{-iE_1t}\textrm{cos}^2\theta_++e^{-iE_2t}\textrm{sin}\varphi_+\textrm{cos}\varphi_++e^{-iE_3t}\textrm{cos}^2\theta_-+e^{-iE_4t}\textrm{sin}\varphi_-\textrm{cos}\varphi_-\\
e^{-iE_1t}\textrm{sin}\theta_+\textrm{cos}\theta_++e^{-iE_2t}\textrm{sin}^2\varphi_++e^{-iE_3t}\textrm{sin}\theta_-\textrm{cos}\theta_-+e^{-iE_4t}\textrm{sin}^2\varphi_-\\
                                \end{array}
                              \right),
\label{Phi1}
\end{aligned}
\end{eqnarray}
which is the detailed result of Eq. (\ref{Phi}).
\end{widetext}
\section{the optimization of electronic observables in the existence of noise}\label{appendixE}
\begin{figure}[b]
	\centering
	\includegraphics[width=0.5\textwidth]{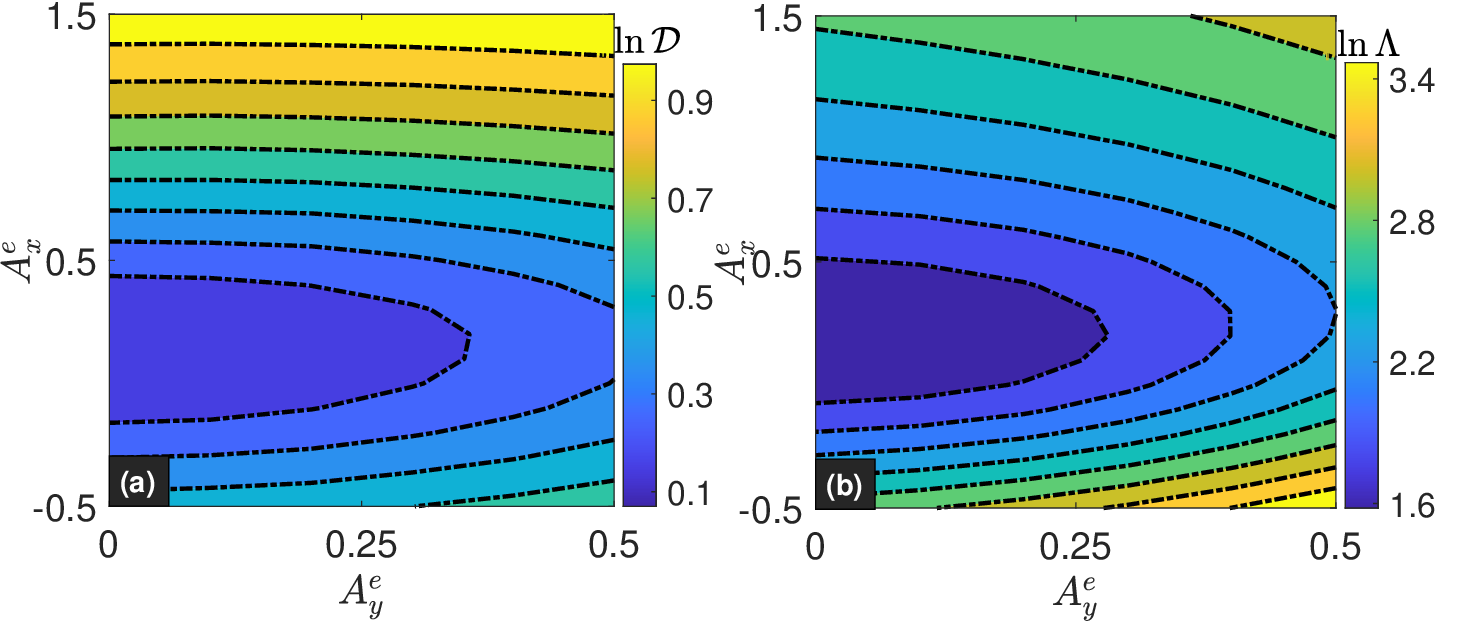}
	\caption {Contour plots for $\ln\mathcal{D}$ and $\ln\Lambda$ of different local observables $A^e$ in the existence of noise inducing dephasing, where $M=\sigma_x$. The parameter is chosen as $\kappa=0.2$, other parameters and the initial state $\tilde{\rho}_0$ are the same as in Fig. \ref{four_com}.}
	\label{reply_3}
\end{figure}
\begin{figure}[b]
	\centering
	\includegraphics[width=0.5\textwidth]{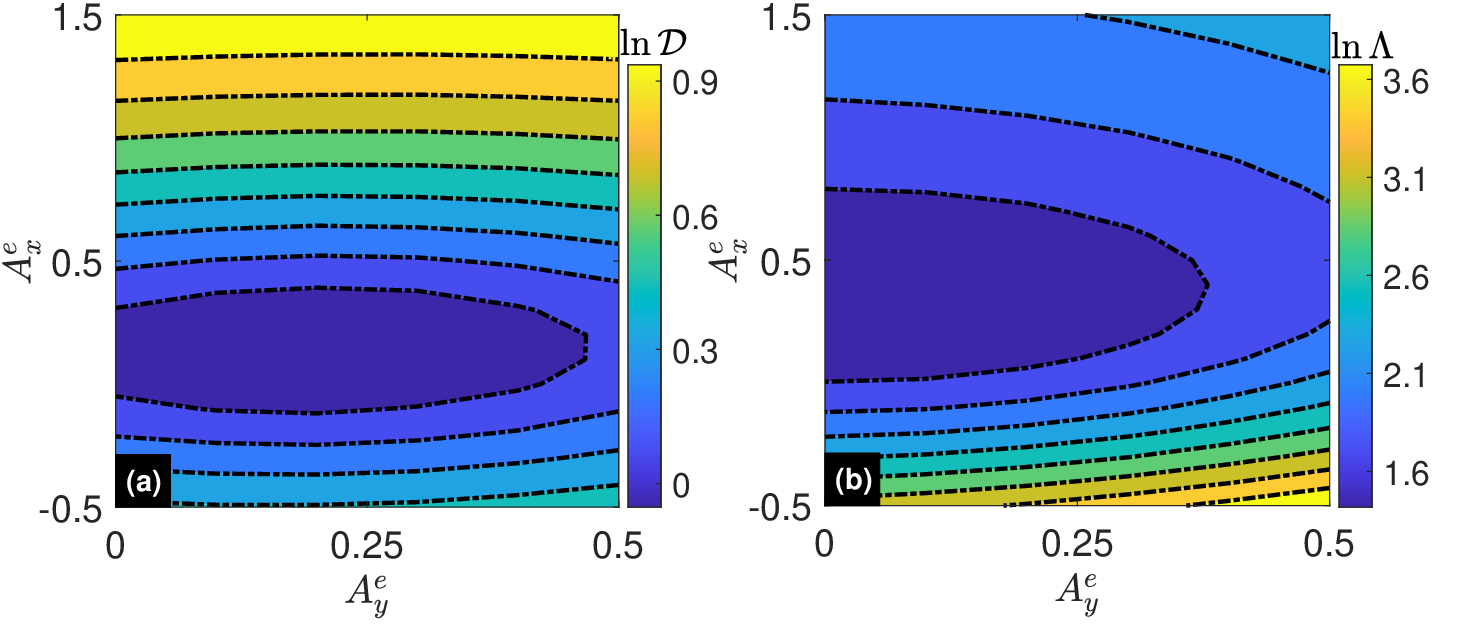}
	\caption {Contour plots for $\ln\Lambda$ and $\ln\mathcal{D}$ of different local observables $A^e$ in the existence of noise inducing dissipation, where $M=\sigma_-$. The parameter is chosen as $\kappa=0.2$, other parameters and the initial state $\tilde{\rho}_0$ are the same as in Fig. \ref{four_com}.}
	\label{reply_4}
\end{figure}
Here, we consider the driving field of the coupled nucleus-electron system is subject to noise, where the noise is regarded as a set of harmonic oscillators with different frequencies. In such a scenario, the dynamic is governed by the master equation with Born-Markov approximation as following \cite{Breuer2007}
\begin{eqnarray}
\begin{aligned}
\dot{\rho}=-i[H_I,\rho]+\frac{\kappa}{2}(2M\rho M^\dagger-\rho M^\dagger M-M^\dagger M\rho),
\label{ME}
\end{aligned}
\end{eqnarray}
where $M$ is the jumping operator with rate $\kappa$ due to the interaction with the noise environment. In the following, we consider two different kinds of noises leading to the dephasing ($M=\sigma_x$) and dissipation ($M=\sigma_-$) process, respectively \cite{Breuer2007}. During the dynamic process, the estimated parameter, nuclear Larmor frequency $\omega_l$ is parameterized into an initial density matrix $\tilde{\rho}_0$. Next, we analyze the effect of the noise on the estimation precision with different electron observables. An arbitrary Hermitian operator of the electron system can be decomposed as Eq. (\ref{Ae}). We give the results of the minimum Euclidean distance $\ln\mathcal{D}$ between the non-optimal observables and the optimal ones in Figs. \ref{reply_3}(a) and \ref{reply_4}(a) for dephasing and dissipation process, respectively. In addition, we optimize the observables of the electronic spin and show the results of $\ln\Lambda$ in Fig. \ref{reply_3}(b) and \ref{reply_4}(b) for different noise process. All the results show that the observable with a shorter minimum Euclidean distance to the optimal ones benefits the estimation precision. Finally, by comparing the results of Fig. \ref{four_com} and the Figs. \ref{reply_3}(b) and \ref{reply_4}(b), we find the minimum $\ln\Lambda$ of Fig. \ref{four_com}(a) is smaller than that in Figs. \ref{reply_3}(b) and \ref{reply_4}(b) which means that the noise is harmful for the estimation.
\end{appendix}

\end{document}